# Band-Order Anomaly at the γ-Al$_2$O$_3$/SrTiO$_3$ Interface Drives the Electron-Mobility Boost

Alla Chikina,[1] Dennis V. Christensen,[2] Vladislav Borisov,[3] Marius-Adrian Husanu,[1,4] Yunzhong Chen,[2] Xiaoqiang Wang,[1] Thorsten Schmitt,[1] Milan Radovic,[1] Naoto Nagaosa,[5] Andrey S. Mishchenko,[5] Roser Valentí,[3] Nini Pryds[2] & Vladimir N. Strocov[1]

[1]*Swiss Light Source, Paul Scherrer Institute, 5232 Villigen-PSI, Switzerland*

[2]*Department of Energy Conversion and Storage, Technical University of Denmark, 2800 Kgs. Lyngby, Denmark*

[3]*Institut für Theoretische Physik, Goethe-Universität Frankfurt am Main, 60438 Frankfurt am Main, Max-von-Laue-Strasse 1, Germany*

[4]*National Institute of Materials Physics, Atomistilor 405A, 077125 Magurele, Romania*

[5]*RIKEN Center for Emergent Matter Science, 2-1 Hirosawa, Wako, Saitama 351-0198, Japan*

# Abstract

Rich functionalities of transition-metal oxides and their interfaces bear an enormous technological potential. Its realization in practical devices requires, however, a significant improvement of yet relatively low electron mobility in oxide materials. Recently, a mobility boost of about two orders of magnitude has been demonstrated at the spinel/perovskite γ-$Al_2O_3$/$SrTiO_3$ interface compared to the paradigm perovskite/perovskite $LaAlO_3$/$SrTiO_3$. We explore the fundamental physics behind this phenomenon from direct measurements of the momentum-resolved electronic structure of this interface using resonant soft-X-ray angle-resolved photoemission. We find an anomaly in orbital ordering of the mobile electrons in γ-$Al_2O_3$/$SrTiO_3$ which depopulates electron states in the top STO layer. This rearrangement of the mobile electron system pushes the electron density away from the interface that reduces its overlap with the interfacial defects and weakens the electron-phonon interaction, both effects contributing to the mobility boost. A crystal-field analysis shows that the band order alters owing to the symmetry breaking between the spinel γ-$Al_2O_3$ and perovskite $SrTiO_3$. The band-order engineering exploiting the fundamental symmetry properties emerges as another route to boost the performance of oxide devices.

The rich interplay of spin, orbital and lattice degrees of freedom in transition metal oxides (TMOs) results in a variety of unconventional phenomena, ranging from superconductivity to ferroelectricity, colossal magnetoresistance and ferromagnetism, which can be further enriched by interfacing these materials. A paradigm example is the two-dimensional electron system (2DES) that can spontaneously form at the interface between two perovskite band insulators LaAlO$_3$ (LAO) and SrTiO$_3$ (STO), for reviews see Refs. [1,2]. This interface combines many intriguing and even mutually exclusive properties such as superconductivity and ferromagnetism.[3] Field-effect tunability of these properties[2,4] allows realization of superconducting field-effect transistors and switchable magnetic states. Whereas the electron concentration ($n_s$) accumulated by the TMO interfaces is typically a couple of orders of magnitude larger compared to the semiconductor interfaces, the electron mobility ($\mu_e$) still stays a few orders of magnitude less.[5] In order to bring the functionality of the TMO interfaces to the level adequate for high-performance practical devices, this fundamental electron transport characteristic needs to be much improved. The limiting factors here include defect scattering, electron-correlation phenomena and, for the STO-based interfaces in particular, strong electron-phonon interaction (EPI) resulting in polaronic nature of the interfacial charge carriers.[6]

Different routes to increase $\mu_e$ of the TMO systems can be envisaged. The most common one is the so-called defect engineering, where the sample preparation procedure is tuned to reduce the concentration of defect scattering sites, including the oxygen vacancies ($V_O$s) or, in the spirit of the semiconductor high-electron-mobility transistors (HEMTs),[7,8] shift their distribution away from the itinerant electrons. An example of such an approach is a modulation-doped heterostructure where a LaMnO$_3$ buffer layer is inserted between amorphous LAO and STO[9] which not only reduces the concentration of $V_O$s on the STO side, but also serves as a spacer to increase the spatial separation between the electrons and the scattering sites.

Recently, the epitaxial heterostructure of spinel γ-Al$_2$O$_3$ (GAO) deposited on perovskite STO was found to exhibit extremely high $\mu_e$ values of up to 140,000 cm$^2$/Vs compared to around 1,000 cm$^2$/Vs in a typical LAO/STO heterostructure.[10] This qualifies GAO/STO as the highest-$\mu_e$ TMO system after the ZnO-based ones.[11] Understanding the fundamental physics behind this $\mu_e$-boost will be extremely important for further progress of the TMO-based devices. A recent photoemission study of GAO/STO supported by first-principles calculations[12] suggested an efficient diffusion of

$V_O$s from the STO bulk to the interface, reducing defect scattering of electrons deeper in STO. However, as the lowest-energy electron states in typical STO-based heterostructures are the $d_{xy}$ ones located in the $V_O$-rich top $TiO_2$ layer,[4,13] it remained unclear why the $d_{xy}$ electrons did not spoil the overall $\mu_e$. Intriguingly, later X-ray linear dichroism (XLD) experiments at the Ti 2$p$ absorption edge[14,15] have suggested a change in the band order in GAO/STO heterostructures, where the $d_{xy}$ states shift above the $d_{xz/yz}$ ones, in contrast to the usual $d_{xz/yz}$>$d_{xy}$ band order at most STO-based interfaces. However, those experiments could not determine the **k**-resolved bandstructure of the 2DES and, most importantly, they probed only the unoccupied states not directly involved in electron transport. Furthermore, the driving force of such band-order anomaly and its connection with the $\mu_e$-boost stayed unclear.

Here, we explore the factors of the $\mu_e$-boost in GAO/STO from direct measurements of the **k**-resolved electronic structure of the buried 2DES, as schematized in Fig. 1, using resonant soft-X-ray angle-resolved photoelectron spectroscopy (ARPES) at the Ti 2$p$ edge. The experiment reveals an anomalous band order, where the $d_{xy}$ band shifts above the $d_{xz/yz}$ ones and depopulates. We reveal that this anomaly shifts the overall electron density away from the $V_O$-rich top STO layer, and the subsequent reduction of the effective defect concentration and the EPI strength, experienced by the mobile electrons, boosts $\mu_e$. A crystal-field (CF) analysis for the interfacial Ti atoms shows that the band-order anomaly is caused by symmetry breaking at the non-isomorph spinel/perovskite interface of GAO/STO. Our analysis puts forward a yet unexplored avenue of the band-order engineering using symmetry breaking to tune properties of the TMO heterostructures towards particular device functionalities.

## Results

The GAO/STO samples were grown by Pulsed Laser Deposition (PLD) under different oxygen pressure resulting in different $n_s$ (see Methods).[16,17] We have investigated a sample having low $n_s$ ~ 3x10$^{13}$ cm$^{-1}$ and $\mu_e$ ~ 12,000 cm$^2$/Vs at 2 K, and another one having high $n_s$ ~ 6x10$^{14}$ cm$^{-2}$ and $\mu_e$ ~100,000 cm$^2$/Vs at 2 K, with such dramatic variations of $n_s$ and $\mu_e$ being typical of STO-based systems. The simultaneous increase of $n_s$ and $\mu_e$ is a hallmark of GAO/STO. Our soft-X-ray ARPES experiments (see Methods) focused on the fundamental electronic structure characteristics – Fermi surface (FS), band dispersions and band order, spectral function – of the interfacial 2DES in these samples compared side-by-side with the paradigm LAO/STO interface irradiated by X-rays to create $V_O$s. Both systems are oxygen-deficient with the $V_O$s concentrated in the top STO layer[18,19] (for more on electronic structure of the localized $V_O$-derived IG states in GAO/STO see the Supporting Information (SI) and the references therein). To access the buried Ti $t_{2g}$-derived 2DES in our samples, we used resonant photoexcitation at the Ti $L_3$-edge at photon energy ($h\nu$) of 460.4 eV. First, we will analyze bandstructure of the 2DES states in terms of electron orbitals.

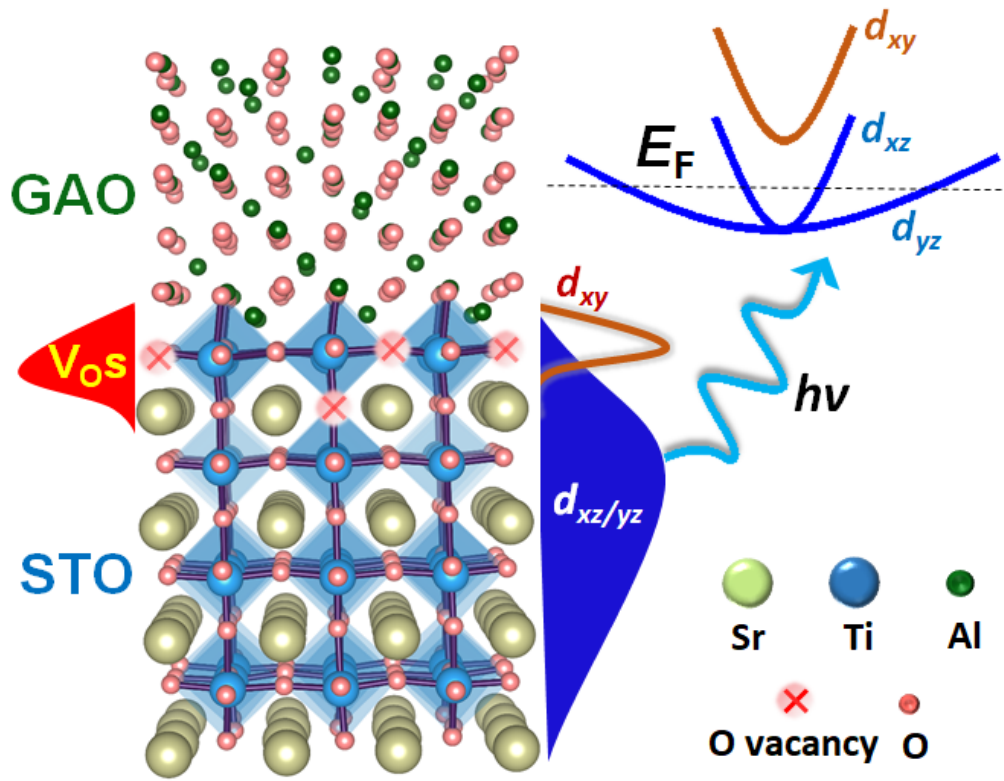

**Fig. 1.** Scheme of the GAO/STO interface probed with **k**-resolved photoemission to directly image electron dispersions. The ARPES experiment finds the band-order anomaly $d_{xy} > d_{xz/yz}$ which promotes spatial separation of the $V_O$s from the 2DES to boost $\mu_e$.

## k-resolved electronic structure: Band-order anomaly

Fig. 2 (a) presents experimental ARPES image of the band dispersions along the ΓX direction of the Brillouin zone for our LAO/STO sample measured with *s*-polarized incident X-rays, selecting the antisymmetric $d_{xy}$ and $d_{yz}$ states. The out-of-plane $d_{yz}$ states show their characteristic flat dispersion, whereas the in-plane $d_{xy}$ one have only a minute cross-section at high excitation energies[20] and is visible mostly in the points where it hybridizes with the $d_{yz}$ states. Fig. 2 (b) represents the ARPES data measured with *p*-polarized X-rays, selecting the symmetric $d_{xz}$ states.[21,22] We note pronounced satellites below the band dispersions, identifying the polaronic nature of the interfacial charge carriers.[6,22]

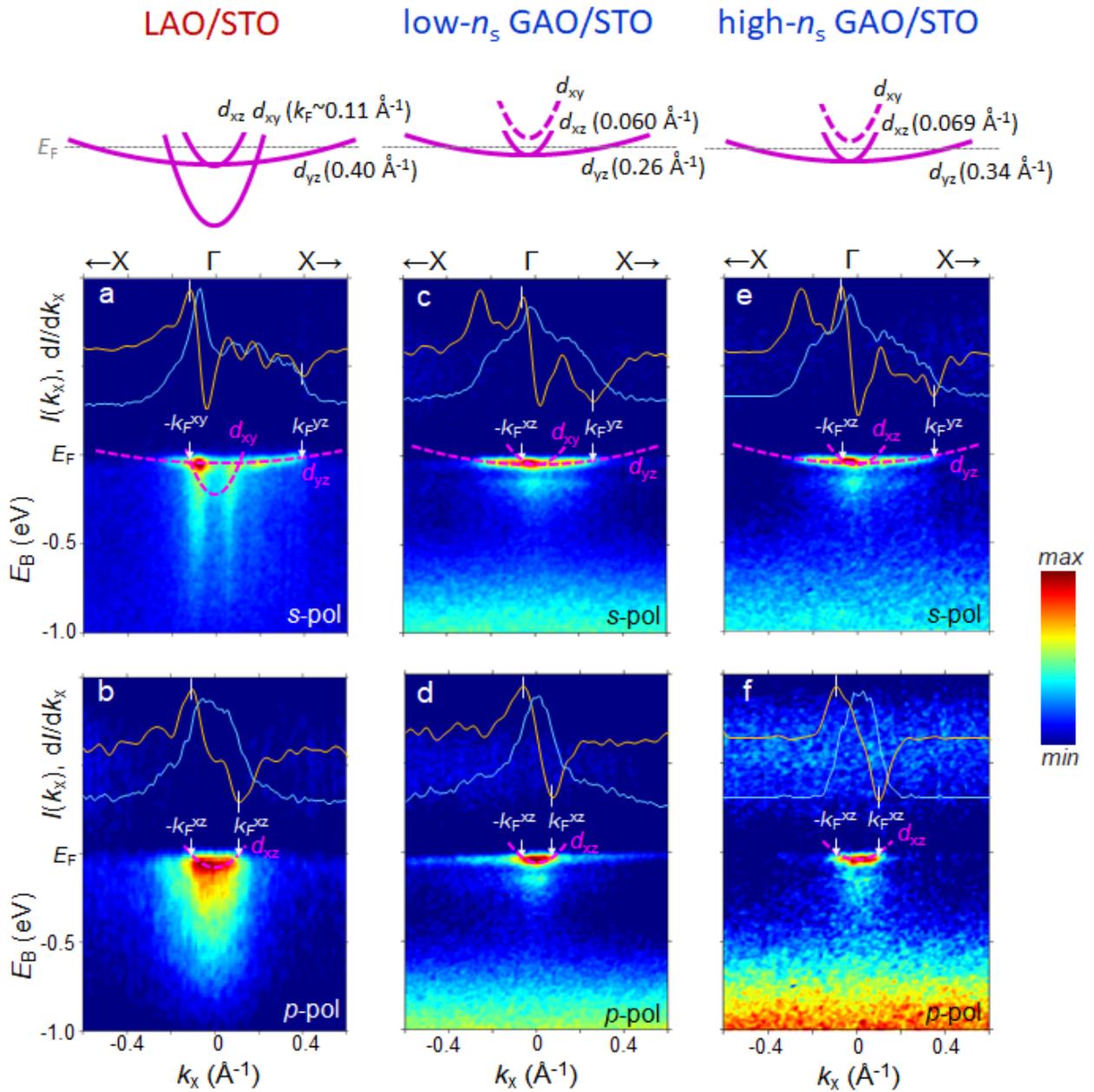

**Fig. 2.** Experimental $E(\mathbf{k})$ images along $\Gamma$X for the oxygen-deficient LAO/STO interface (a,b) and low- and high-$n_s$ GAO/STO interfaces (c,d and e,f, respectively) measured at $h\nu = 460.4$ eV. For LAO/STO, $s$-polarized X-rays select the $d_{xy}$- and $d_{yz}$ bands (top panels) and $p$-polarized X-rays the $d_{xz}$ bands (bottom panels), and for GAO/STO these selection rules relax. The MDCs at $E_F$ (cyan curves) and their gradients (yellow) are shown on top of each panel, with extremes of the latter (dashes) locating the $k_F$ values (arrows). The sketches on top of the figure schematize the band order for the three samples, where GAO/STO shows the anomalous $d_{xy} > d_{xz/yz}$ band order with complete depopulation of the $d_{xy}$ band.

Importantly, the experimental bandstructure of LAO/STO follows the $d_{xy}<d_{xz/yz}$ band order, with the $d_{xy}$ band minimum shifted by ~200 meV below that of $d_{xz/yz}$.[19] Schematized on top of Fig. 2 (a), such pattern confirms the theoretical picture where the CF in bulk STO lowers the energy of Ti $t_{2g}$ bands relative to the $e_g$ ones, and interfacing STO with LAO further splits the former into the deeper $d_{xy}$-derived band and more shallow $d_{xz}/d_{yz}$ ones.[21–23] The $d_{xy}$ states are located close to the interface, while the $d_{xz}/d_{yz}$ ones extend into the STO bulk over a region of up to a few tens nanometers, depending on the exact shape of the interfacial quantum well $V(\mathbf{r})$.[4,6,13] The $d_{xy}<d_{xz/yz}$ band order observed in LAO/STO is actually universal across a wide range of systems based on $TiO_2$-terminated STO almost independently of the growth conditions and oxygen deficiency. It starts from the paradigm LAO/STO interface[21] and survives under amorphous overlayers of LAO,[24,25] Si[26] and a variety of metals[27–29] on STO as well as for bare STO surfaces prepared under various conditions[30,31] and even reconstructed by sputtering/annealing.[32] This band order can though be tuned by applying pressure[33] or changing crystallographic orientations,[34] although the latter also changes the $d_{xy}/d_{xz}/d_{yz}$ orbital orientation relative to the surface or interface.

For GAO/STO, the experimental bandstructure of the $d_{xy}$- and $d_{yz}$ states shown in Fig. 2 (c) and (e) for the low- and high-$n_s$ samples, respectively, appears quite differently from LAO/STO. First of all, the proximity of GAO breaks the fourfold symmetry of STO due to the interfacial Ti cations which neighbor the tetrahedrally coordinated Al cations. Due to the concomitant relaxation of the above selection rules, at the *s*-polarization we observe certain weight of the $d_{xz}$ band on top of the $d_{yz}$ one (c,e), and at the *p*-polarization certain weight of the $d_{yz}$ band on top of the $d_{xz}$ one (d,f). Most importantly, the $d_{xy}$-$d_{yz}$ hybridization points characteristic of LAO/STO disappear, indicating the absence of the $d_{xy}$ band which has shifted above $E_F$ and depopulated. This bandstructure pattern, schematized on top of the figure, identifies an anomalous $d_{xy}>d_{xz/yz}$ band order as opposed to the universal $d_{xy}<d_{xz/yz}$ one represented by the above LAO/STO. The intensity enhancement at the bottom of the GAO/STO images is due to the $V_O$-derived IG states (see the SI). We note that the band-order is associated with a small change of the effective mass of the $d_{xz/yz}$ bands (see the SI) which, as we discuss later, can be related to the reduction of the EPI.

Therefore, our ARPES results reveal that electron transport in GAO/STO is exclusively due to the $d_{xz/yz}$-electrons. We note that in many cases one can deduce the partial band occupancy from non-linearity of the Hall coefficient as a function of magnetic field. In the case of high-mobility GAO/STO samples, however, the Hall coefficient shows a very pronounced anomalous Hall effect stemming from the magnetism, and the concomitant non-linearity hides possible multiband effects. ARPES, in turn, clearly resolves the orbital character of the contributions to the total $n_s$.

We note that our ARPES findings of the anomalous band order in GAO/STO are consistent with the recent XLD experiments[14,15] which are though relevant for the band order in the unoccupied states not directly related to the electron transport. Ionic liquid gating experiments[17] suggested the persistence of the universal band order in GAO/STO, but these experiments probe the band structure only indirectly and may be affected by electrochemical reactions.

The Fermi momenta ($\mathbf{k}_F$) of the experimental bands, reflecting their population, were determined from the momentum-distribution curves (MDCs) of the ARPES intensity $I(k_x)$ integrated over the whole occupied energy width ($W$) of the bands (cyan lines in the bandstructure panels of Fig. 2). The extremes of their gradient $dI/dk_x$ (yellow lines) define the $\mathbf{k}_F$ values indicated at the band schemes on top of Fig. 2. We have utilized this gradient method originally proposed by Straub et al.[35] because in our case, where $W$ is of the order of the experimental $\Delta E$, the conventional determination of $\mathbf{k}_F$ from the maxima of the MDCs at $E_F$ becomes obviously irrelevant. Assigning the $dI/dk_x$ extremes to particular $\mathbf{k}_F$, we should keep in mind that at the negative-$k_x$ side of the $d_{yz}$ bands and positive-$k_x$ side of the $d_{xy}$ ones the spectral intensity vanishes due to photoemission matrix elements. Inspecting the experimental $\mathbf{k}_F$, we observe that not only does the $d_{xy}$ band depopulate when going from LAO/STO to GAO/STO, but also the $d_{xz/yz}$ bands reduce their population; these bands increase their population between the low- to high-$n_s$ GAO/STO samples. The relation of these observations with $n_s$ and $\mu_e$ of the 2DES will be discussed later.

## Symmetry-breaking origin of the band-order anomaly

In order to get more insight into the origin of the band-order anomaly at the spinel-perovskite GAO/STO interface, we have theoretically analyzed the CF splitting effects in the ion-core regions of the interfacial Ti atoms. The CF potential was modeled for different atomic environments depicted in Fig. 3 (a,b). The point-charge approximation is used to calculate the spatial distribution of the electrostatic potential of $N$ ions with nominal charges $Q_i$ and coordinates $r_i$ acting on electrons:

$$V_{CF}(\mathbf{r}) = -\frac{1}{4\pi\varepsilon_0}\sum_{i=1}^{N}\frac{Q_i}{|\mathbf{r}-\mathbf{r}_i|}$$

The 3$d$ states are approximated by the atomic orbitals with $n=3$ and $l=2$, which are used for evaluating the CF matrix:

$$H_{CF} = <nlm_1|V_{CF}(r) - V_{CF}(0)|nlm_2>.$$

Here, $|nlm>$ stands for spherical harmonics, the values of $m_1$ and $m_2$ are between $-2$ and $+2$ and correspond to $d_{z^2}$, $d_{x^2-y^2}$, $d_{xy}$, $d_{xz}$ and $d_{yz}$ orbitals. Diagonalization of the CF matrix $H_{CF}$ gives information on the band ordering at the Γ-point for the selected atomic structure.

For the cubic STO atomic environment, Fig. 3 (a), the $d_{xy}$ vs $d_{xz/yz}$ orbitals are degenerate. The degeneracy can be lifted, for example, by symmetry breaking when interfacing STO to other TMOs such as LAO, or by oxygen octahedra tilting. The latter, illustrated in Fig. 3 (c) where the tilting around the x- or y-axes pushes the $d_{yz}$ or $d_{xz}$ states above the $d_{xy}$ one, is typical of TMOs and constitutes one of the contributions to the interfacial potential in oxide heterostructures. The symmetry breaking and oxygen-octahedra tilting, both lead to a shift of the $d_{xz/yz}$ orbitals above the $d_{xy}$ one, Fig. 3 (c), which was, indeed commonly observed as the universal band order in numerous experimental and theoretical studies of STO-based interfaces.[19,21,36,37]

For the spinel-perovskite environment, Fig. 3(b), we find that the CF energy levels are dramatically rearranged due to the proximity of positively charged Al cations, which can be either octahedrally or tetrahedrally coordinated by oxygen. The particular coordination is crucial for the energy position of the IG states in GAO/STO formed by the $V_O$s, but does not qualitatively change the anomalous band order of the metallic states which stays inverse compared to LAO/STO. Our CF calculations suggest that the lowest energy level becomes dominated by the $d_{xz}/d_{yz}$ states that are mixed in 1:1 ratio while the next higher energy levels are mostly of the $+d_{xy}$ character, Fig. 3 (d). We emphasize that this feature is robust with respect to whether the Al vacancies, intrinsically present in GAO, are located in the Al-layer close to STO or elsewhere in GAO. In the studied example, the $V_O$s increase the energy separation between the $d_{xy}$ and $d_{xz/dyz}$ orbitals slightly but do not change this picture qualitatively, simply providing electrons filling energy levels. Therefore, our analysis suggests that the band-order anomaly in GAO/STO originates from the different nature of the CF in the ion-core regions of the interfacial Ti atoms at this spinel-perovskite interface compared to the perovskite-perovskite LAO/STO one.

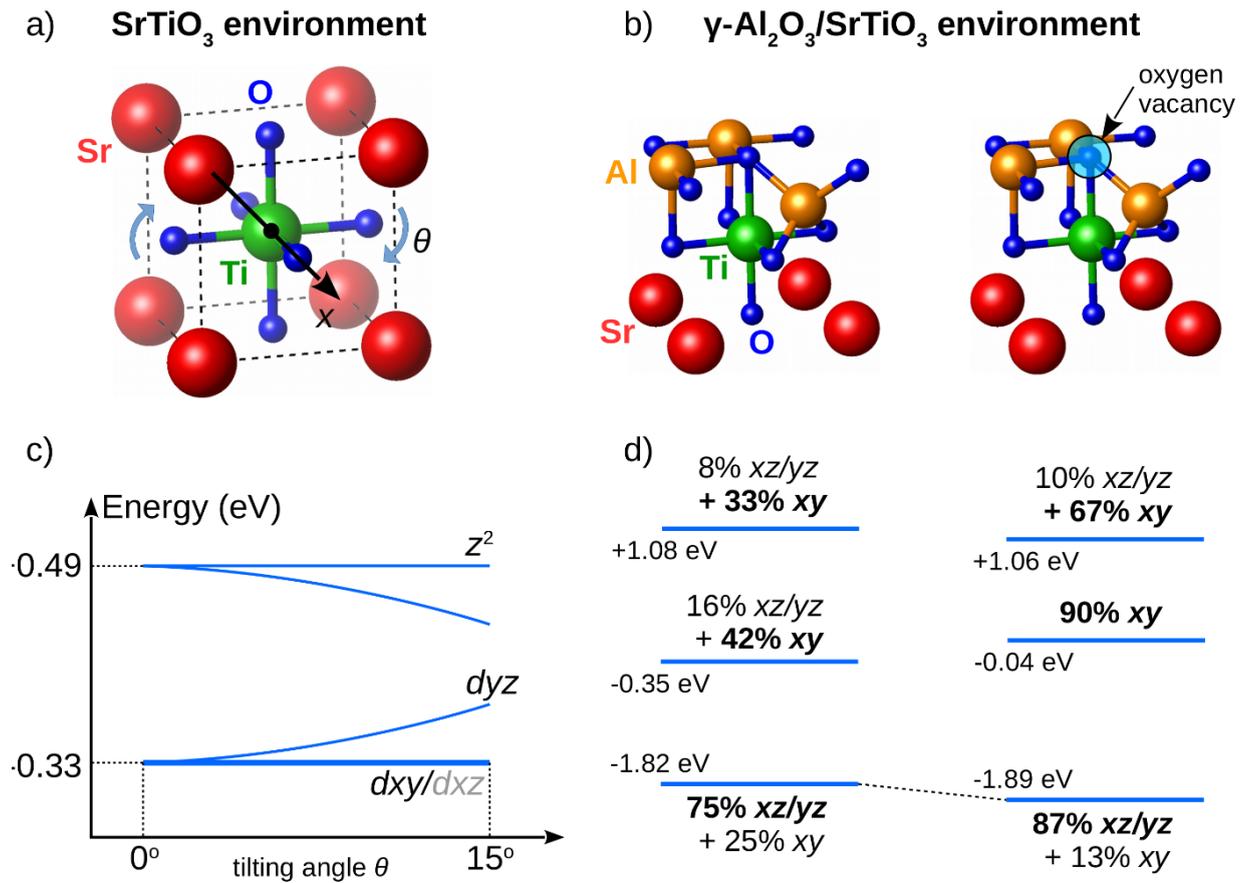

**Fig. 3.** CF analysis for different atomic environments. a) Structural sketch of bulk perovskite STO. b) Structural sketch of one possible termination of the spinel-perovskite GAO/STO interface. c) CF energy levels for bulk STO as a function of the oxygen octahedra tilting angle. d) Lowest CF energy levels for defect-free and oxygen-deficient GAO/STO environments, with the orbital character and weights indicated (the dominant weight in bold).

## Spectral function: Electron-phonon interactions and disorder

We will now analyze another important ingredient of the electronic properties, the EPI, which is embedded in the spectral function $A(\omega,\mathbf{k})$ represented by the energy-distribution curves

(EDCs) of the ARPES intensity in Fig. 2. Similarly to the previous analysis for the stoichiometric LAO/STO,[6,22] we will use $A(\omega)$ obtained by integration of $A(\omega,\mathbf{k})$ over the whole $\pm k_F$ interval of the $d_{yz}$ bands. This approach allows full inclusion of the bosonic spectral satellites whose extension in **k**-space may be narrower than the quasiparticle (QP) bands.[38] The experimental $A(\omega)$ for our oxygen-deficient LAO/STO and both GAO/STO samples (where we subtracted the IG state weight protruding into the 2DES energy region) are presented in Fig. 4 (a). Their pronounced peak-dip-hump lineshape is a hallmark of the strong EPI in STO-based systems, fundamentally reducing $\mu_e$.[6,22] We recall that whereas for LAO/STO the $A(\omega,\mathbf{k})$ reflects both $d_{xy}$ and $d_{xz/yz}$ states, for GAO/STO it reflects only the latter.

For LAO/STO the hump is structureless and extends over a wide energy range, suggesting a continuum of phonon modes involved in the EPI. For GAO/STO, in contrast, the hump is a narrow peak separated from the QP one by ~100 meV which corresponds to the LO3-phonon energy in STO.[32] The increase of the QP residual weight $Z_0$ from ~31% in LAO/STO to ~45% in GAO/STO signals a decrease of the EPI strength in the latter. Expectedly, this increase is associated with a notable reduction of the effective mass of the $d_{xz/yz}$ bands (see the SI) although a limited accuracy of its experimental values impedes a quantitative analysis on this point.

Finally, we will analyze the disorder experienced by the charge carriers. Fig. 4 (b) presents the QP peak in vicinity of $E_F$ whose $A(\omega,\mathbf{k}=\mathbf{k}_F)$ was obtained as EDCs integrated within $\pm 0.1$ Å$^{-1}$ around $k_F$ of the $d_{yz}$ bands. In this case the band-dispersion effects are minimized and, within the Fermi-liquid paradigm, the spectral broadening due to electron-electron interaction vanishes at $E_F$. Therefore, the QP-peak broadening ($\Delta E$) which, neglecting the experimental energy resolution, is inversely proportional to the coherence length of the charge carriers, reflects the effective disorder they experience. In comparison with the LAO/STO's value $\Delta E \sim 62$ meV, for the GAO/STO samples we observe $\Delta E \sim 50$ meV, which is close to the resolution limit ~40 meV of our experiment. The reduced $\Delta E$ expresses the smaller effective disorder experienced by the $d_{yz}$-derived charge.

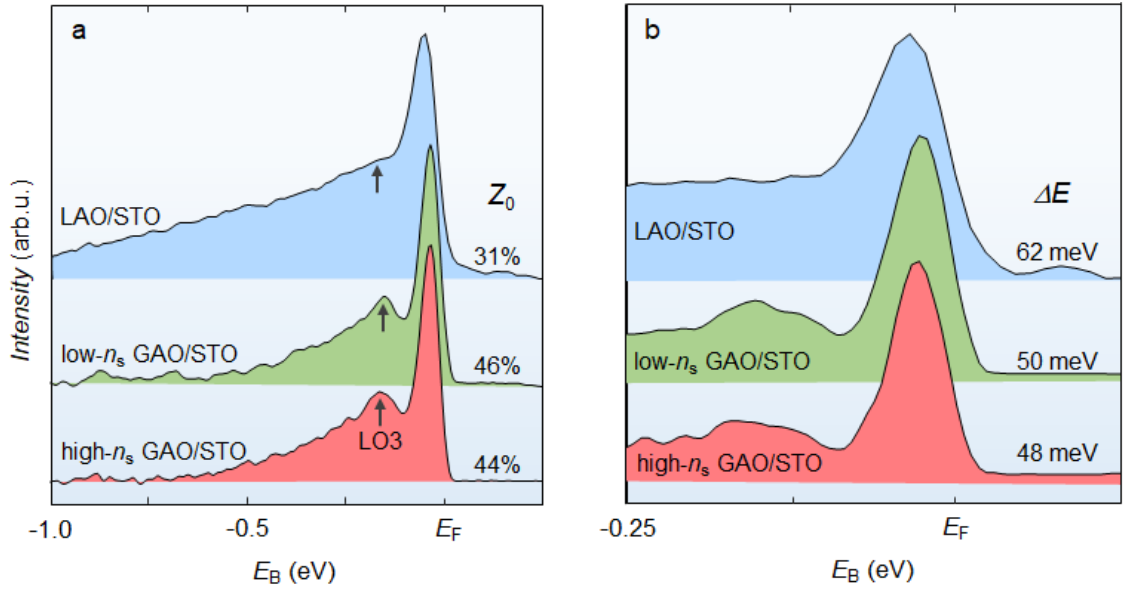

**Fig. 4.** Experimental $A(\omega,\mathbf{k})$ for oxygen-deficient LAO/STO and two GAO/STO samples represented by EDCs of the ARPES intensity from Fig. 2 integrated over the (a) whole $\pm k_F$ interval and (b) $k_F \pm 0.1$ Å$^{-1}$ interval. Compared to LAO/STO, in both GAO/STO samples the QP peak shows larger spectral weight $Z_0$ and smaller energy broadening $\Delta E$ (values on the plots) which identify weaker EPI and smaller effective disorder, respectively.

# Discussion

### Effect of the band order on $\mu_e$

The central question is now how does the observed band-order anomaly affect $\mu_e$? Fig. 1 sketches the distribution of the $V_O$s relative to the depopulated $d_{xy}$ and populated $d_{xz/yz}$ electron density in GAO/STO. The $V_O$s effectively diffuse to the top STO layer, reducing their concentration and thereby the associated defect scattering in the deeper STO region.[12] This reduction is confirmed by our analysis of the QP peak width from the $d_{xz/yz}$-electrons extending into this region, which has evidenced a small effective disorder experienced by these electrons compared to LAO/STO. Located in the defect-free region, the $d_{xz/yz}$-electrons deliver high $\mu_e^{xz/yz}$, whereas the $d_{xy}$-states, located in the defect-rich top layer,[4,13,21] are depopulated and thus do not poison the overall $\mu_e$. This interplay between the spatial distribution and population of the electron states is the main driving force behind the boost of $\mu_e$ in GAO/STO. Obviously, the boost critically depends on the spatial distribution of the $V_O$s and other defect sites, highly sensitive to the sample growth and annealing protocol.[12]

The $d_{xy}$>$d_{xz/yz}$ band order will affect virtually all the properties of the GAO/STO besides $\mu_e$. For example, it should reverse the sign of the gating voltage corresponding to the Lifshitz transition, where the crossover from the one-band to multiband regime causes dramatic changes in physical properties of the STO-based heterostructures including the superconducting transition temperature, electron pairing strength, and spin-orbit coupling.[39] Moreover, our ARPES findings may also have implications for the mechanism of superconductivity in GAO/STO, as this phenomenon in the STO-based systems is usually observed in the multiband conduction regime.[40] Both points still await experimental verification though.

## Depth extension of the 2DES

The connection of the ARPES data to the carrier density is established by the Luttinger theorem.[41] For the 2D systems, it states that $n_s$ is equal to the sum of the partial Luttinger counts $2\int_{FS} \frac{d^2k}{(2\pi)^2}$ over all sheets constituting the FS. However, the application of this theorem to our case is not straightforward even knowing the experimental FS of GAO/STO (see the SI). Firstly, a typical feature of the STO-based systems is the electronic phase separation where conducting regions co-exist with insulating ones.[18,42,43] While experimental FS reflects the local electron density in the conducting areas, the integral $n_s$ can be smaller. Secondly, if the interfacial $V(\mathbf{r})$ is sufficiently long-range, it hosts a ladder of quantum-well (QW) states,[44] and the Luttinger count should include the FS sheets through all of them. However, whereas the first QW state is located close to the interface, the higher-order ones are shifted deeper into the STO bulk[45] where they can escape detection within the probing depth of the ARPES experiment while contributing to the total $n_s$. Because of the missing higher-order FS sheets, the Luttinger count of the FS apparent in ARPES may underestimate the $n_s$ measured in transport.

Intriguingly, the experimental $\mathbf{k}_F$ values found by ARPES for GAO/STO show an increase of the corresponding Luttinger counts by only ~15% between the low- and high-$n_s$ samples, which is much less compared to the increase of transport $n_s$ by a factor of ~20. This discrepancy may manifest the two effects discussed above. First, the lateral increase of the conducting phase between the low- and high-$n_s$ samples, estimated from intensity of the 2DES signal as described in Ref. [18], measures ~70%. The remaining contribution to the $n_s$ increase of more than an order of magnitude should then manifest a long-range $V(\mathbf{r})$ in the high-$n_s$ sample, which supports an electron-density accumulation in higher-order $d_{xz/yz}$ derived QW states located in STO beyond the top 2-3 u.c. reached by ARPES. This situation can be viewed as a transformation of the 2DES into a quasi-three-dimensional system expanding deep into $V_O$-free STO. In this case the increase of $n_s$ should be associated with an increase of $\mu_e$, and it indeed increases by a factor of ~8 between our samples. This physical picture resolves the long-standing puzzle why GAO/STO denies the 'small $n_s$ - large $\mu_e$' paradigm often advocated for oxide interfaces, although other cases have been reported where this paradigm was violated by similar dimensionality transformations.[46]

## Effects of the EPI

The weakening of the EPI, expressed by the $Z_0$ values in Fig. 4 (a), is another aspect of GAO/STO contributing to the $\mu_e$-boost. For many systems such weakening is caused by screening of the EPI by mobile electrons. In our case, however, ARPES shows that at least near the interface the $d_{xz/yz}$ derived electron density in GAO/STO is smaller than the $d_{xz/yz}$ (and all the more $d_{xy}+d_{xz/yz}$) density in our LAO/STO. In fact, the EPI may be amplified by trapping (localization) of charge carriers on defects[47] or *via* more exotic Anderson localization.[48] This lowers the crystal symmetry, allowing otherwise prohibited phonon modes and thus additional EPI channels. The involvement of such localization-enhanced EPI in our case is evident from the phonon-mode continuum displayed by LAO/STO's $A(\omega)$ as opposed to the single mode in GAO/STO. Then the reduction of the effective defect concentration in GAO/STO, expressed by the smaller disorder seen in the QP-peak width, Fig. 4 (b), explains the reduction of the EPI. This effect traces back, again, to the band-order anomaly that shifts the total 2DES into the $V_O$-free bulk of STO. The small but significant decrease of EPI in the high-$n_s$ GAO/STO sample compared to the low-$n_s$ one should be a combined effect of smaller effective disorder due to the deeper expansion of the 2DES into STO, and of larger $n_s$ screening the EPI.

The observed weakening of EPI in GAO/STO propagates into $\mu_e$ not only by polaronic renormalization of the electron effective mass, but also by mediating electron scattering on defects, which can somehow be viewed as frozen phonons. Therefore, the defects affect $\mu_e$ not only through their concentration but also through the EPI-mediated electron-defect scattering strength which, in turn, increases with the defect concentration. This cumulative effect results in a steep non-linear change of $\mu_e$ as a function of the latter, predicted within the mean-field approximation already four decades ago[49] and later reproduced by approximation-free Diagrammatic Monte Carlo calculations.[50–52] This effect is likely to be at play to boost $\mu_e$ with decrease of the effective $V_O$ concentration experienced by the $d_{xz/yz}$ electrons in GAO/STO thanks to the band-order anomaly. Indeed, this decrease is evident from our analysis of the QP peak width in Fig. 4 (b). The tendency of $V_O$s to cluster[18] can also affect $\mu_e$.

## Conclusions and outlook

To summarize our results, we have found that the **k**-resolved electronic structure of the 2DES at the GAO/STO interface stands out of all other known STO-based interfaces in terms of the band order. The upward shift of the $d_{xy}$-states with respect to the $d_{xz}/d_{yz}$ ones, forming the anomalous $d_{xy}>d_{xz/yz}$ band order and depopulating the $d_{xy}$ band, is attributed to a change in the CF configuration around the Ti ions as the spinel/perovskite GAO/STO interface breaks the perovskite-lattice symmetry. The band-order affects virtually all physical properties of GAO/STO, and in the first place boosts $\mu_e$ of the interfacial 2DES which shifts away from the defect-rich interface towards the STO bulk.

The $\mu_e$-boost at the GAO/STO interface appears as a multifaceted phenomenon, identifying three routes to design high-mobility oxide devices in general: (1) The static defects, in particular the $V_O$s, affect $\mu_e$ through low-temperature electron scattering. With inspiration from the semiconductor HEMTs, this scattering can be reduced by shifting the depth profile of $V_O$s away from that of the 2DES to minimize their overlap. In GAO/STO this is achieved by the accumulation of the $V_O$s at the top $TiO_2$ layer of STO away from the 2DES located deeper in STO[12]. The crucial role of this *defect-engineering* route for GAO/STO can be assessed by varying the growth protocol and thus distribution of $V_O$s;[53] (2) This work suggests the second route, *band-order engineering*. Complementary to the defect engineering, this route allows control over the 2DES depth profile in order to shift it away from the $V_O$s. In GAO/STO this is achieved with the $d_{xy}>d_{xz/yz}$ band order, which depopulates the $d_{xy}$ states and leaves only the $d_{xz/yz}$-electrons, located deeper in the STO bulk compared to the $d_{xy}$ ones in the defect-rich top layer. Moreover, the increase of $n_s$ in GAO/STO expands the 2DES into the STO bulk and thus further increases $\mu_e$. The band order can be manipulated through the CF driven by the interface-symmetry breaking and, in addition, tuned by doping such as substituting $Al^{3+}$ ions in GAO with those of a dissimilar valence state; (3) The third route to boost $\mu_e$ is the *EPI-engineering* route that invokes the EPI-mediated electron interaction with the defects. As the EPI itself is enhanced by electron trapping on the defects, the resulting $\mu_e$ shows a steep non-linear dependence of their effective concentration, in our case that of the $V_O$s over the 2DES spatial extension. In this context the EPI-engineering is intimately connected with the defect and band-order engineering. The EPI can also be tuned through $n_s$ varied, for example, by electrostatic gating.[54] Thereby, the GAO/STO interface teaches us a triad of interconnected routes – defect-, band-order and EPI-engineering – towards realization of a high-mobility 'oxide HEMT'. None of them alone but only the whole triad, and perhaps assisted by other yet unknown mechanisms, can exhaust the entire $\mu_e$-boost in GAO/STO of more than two orders compared to the paradigm LAO/STO heterostructures.

In a broader perspective, the crucial role of the band-order engineering extends from merely electron transport towards the whole world of other exotic properties of the oxide systems including gate-tunable superconductivity, Lifshitz transition, electron pairing without superconductivity, spin-orbit coupling, magnetic ordering, *etc*. For example, the $d_{xy}>d_{xz/yz}$ band order observed in GAO/STO should revert the gate-voltage sign of the Lifshitz transition as well as switch the superconductivity from the multiband to one-band regime. Our work puts forward a prospect to change the band order through an appropriate lattice-symmetry breaking, and tuning the CF splitting by dopants of a dissimilar valence state or by (pseudo) epitaxial growth of other members of the large spinel crystal family. Our findings motivate searching of other oxide interfacial pairs, where the $d_{xy}$ bands totally depopulate to maximize $\mu_e$, using modern big-data computational methods.

# Methods

## Sample fabrication and characterization

The samples were manufactured by pulsed laser deposition (PLD) on STO substrates. The reference LAO/STO samples were grown in slightly oxygen-deficient conditions and annealed in oxygen *ex-situ*, following the protocol described in Refs. [18,19]. The GAO/STO samples were manufactured using the protocol described elsewhere[16] under the oxygen partial pressure varied from 4 to $8 \times 10^{-6}$ mbar in order to change the transport properties through the $V_O$ concentration. The two investigated GAO/STO samples were metallic with low $n_s$ ~ 3x10$^{13}$ cm$^{-2}$ and high $n_s$ ~ 6x10$^{14}$ cm$^{-2}$ at room temperature with the residual resistance ratio $R_s$(300K)/$R_s$(2K) of ~1100 and 9300, respectively. The high residual resistance ratios are the hallmarks of samples with high low-temperature $\mu_e$,[16] which was ~12,000 cm$^2$/Vs and 100,000 cm$^2$/Vs at 2 K for the two samples.

## ARPES experiment

The experiments were performed at the soft-X-ray ARPES facility[55] installed at the ADRESS beamline[56] of the Swiss Light Source, Paul Scherrer Institute, Switzerland. Variable X-ray polarization delivered by this beamline allowed symmetry analysis of the electron states. The FS maps were recorded at a combined (beamline plus analyzer PHOIBOS-150) energy resolution of ~60 meV, and $E(\mathbf{k})$ maps at ~40 meV. Angular resolution of the analyzer was ~0.1°. The sample was cooled down to ~12K in order to quench relaxation of $\mathbf{k}$-conservation due to thermal motion of the atoms. The photon flux was ~10$^{13}$ photons/s and focused into a spot of 30 x 75 µm$^2$ on the sample surface. Other relevant experimental details are reported elsewhere.[6,19]

During the ARPES data acquisition, the strength of the 2DES signal from our LAO/STO and GAO/STO samples gradually increased because the V$_O$s generated under X-ray irradiation injected mobile electrons into the 2DES.[18,19] However, the observed band dispersions and population stayed constant, following the electronic phase separation scenario typical of the STO-based systems.[18,20,42,43] The irradiation-induced increase of the 2DES signal from the high-$n_s$ GAO/STO sample was smaller compared to the low-$n_s$ one. All reported ARPES data were acquired after an irradiation time of more than 1 hour to insure saturation. The experimental resonant-photoemission intensity maps as a function of $h\nu$ through the Ti $L$-edge for our LAO/STO and low-$n_s$ GAO/STO samples are presented in the SI. Intensity of the V$_O$-derived IG states at a binding energy around -1.2 eV and of the 2DES states near $E_F$ sharply varied with $h\nu$. The data in the main text have been acquired at the $L_3$-resonance at $h\nu$ = 460.4 eV, where the $d_{xy}$-to-$d_{xz/yz}$ intensity ratio enhances compared to the $L_2$-resonance.[6,19] The corresponding FS maps of GAO/STO measured at different X-ray polarizations are presented in the SI. Besides the reduced FS area, their notable difference to the FS maps of LAO/STO is notable streaks of intensity stretching between the Γ points. This peculiarity might be related to linear clusters of the interfacial V$_O$s or, although scanning-SQUID measurements on similarly prepared samples have found micron-size domains, the existence of domains whose size would be below the coherence length of the ARPES experiment (for details see the SI).

## Associated Content

The SI includes (1) Ti *L*-edge resonant photoemission data, (2) effective-mass analysis, (3) DFT calculations, and (4) experimental Fermi surface. The SI is available online.

## Author information

### Corresponding authors

**Vladimir N. Strocov** – *Swiss Light Source, Paul Scherrer Institute, 5232 Villigen-PSI, Switzerland*; Phone: +41-56-310-5311; orcid.org/0000-0002-1147-8486; Email: vladimir.strocov@psi.ch

**Nini Pryds** – *Department of Energy Conversion and Storage, Technical University of Denmark, 2800 Kgs. Lyngby, Denmark*; orcid.org/0000-0002-5718-7924; Phone: +45-46-77-5752; Email: nipr@dtu.dk

### Author contributions

A.C., D.C., Y.C., M.A.H., M.R., N.P. and V.N.S. performed the ARPES experiment assisted by X. W. and T.S. V.N.S, A.C., and M.A.H. processed the ARPES data. D.C. and Y.Z.C. supported by N.P. fabricated the samples and performed their transport characterization. V.N.S. and N.P. conceived the idea of the band-order effect on electron mobility, and A.S.M. supported by N.N. elaborated the physics of EPI. V.B. supported by R.V. performed the CF calculations. V.N.S. wrote the manuscript with contributions from A.C., D.C., V.B. and A.M. All authors discussed the results, interpretations, and scientific concepts.

### Notes

The authors declare no competing financial or non-financial interests.

# Acknowledgements

We thank L. Nue for skillful technical support of the SX-ARPES experiments. A.C. acknowledges funding from the Swiss National Science Foundation under Grant no. 200021-165529. M.-A.H. was supported by the Swiss Excellence Scholarship grant ESKAS-no. 2015.0257 and the Romanian UEFISCDI Agency under Contracts No. 475 PN-III-P4-ID-PCCF2016-0047. A.S.M. and N.N. acknowledge support of JST CREST Grant Number JPMJCR1874, Japan. V.B. and R.V. were supported by DFG Sonderforschungsbereich TRR 49 and by the computer center of Goethe University Frankfurt.

# Band-Order Anomaly at the γ-Al$_2$O$_3$/SrTiO$_3$ Interface Drives the Electron-Mobility Boost (Supporting Information)

Alla Chikina,[1] Dennis Valbjørn Christensen,[2] Vladislav Borisov,[3] Marius-Adrian Husanu,[1,4] Yunzhong Chen,[2] Xiaoqiang Wang,[1] Thorsten Schmitt,[1] Milan Radovic,[1] Naoto Nagaosa,[5] Andrey S. Mishchenko,[5] Roser Valentí,[3] Nini Pryds[2] & Vladimir N. Strocov[1]

[1]*Swiss Light Source, Paul Scherrer Institute, 5232 Villigen-PSI, Switzerland*

[2]*Department of Energy Conversion and Storage, Technical University of Denmark, 2800 Kgs. Lyngby, Denmark*

[3]*Institut für Theoretische Physik, Goethe-Universität Frankfurt am Main, 60438 Frankfurt am Main, Max-von-Laue-Strasse 1, Germany*

[4]*National Institute of Materials Physics, Atomistilor 405A, 077125 Magurele, Romania*

[5]*RIKEN Center for Emergent Matter Science, 2-1 Hirosawa, Wako, Saitama 351-0198, Japan*

# Ti *L*-edge resonant photoemission: 2DES *vs* IG states

The Ti *L*-edge resonant photoemission (ResPE) maps for our oxygen-deficient LAO/STO and low-$n_s$ GAO/STO are presented in Fig. S1. The maps were acquired at a saturation dose of X-ray irradiation. The ARPES intensity was integrated within $\pm k_F^{yz}$. The maps show the valence band (marked VB), IG states (IG) and mobile 2DES states (2DES) whose spectral response dramatically varies as a function of excitation energy. The VB signal, resonating at the $Ti^{4+}$ $t_{2g}$ and $e_g$ peaks of X-ray absorption, represents the oxygen-derived states hybridized with Ti in the STO bulk. The VO-derived IG states resonate at the $Ti^{3+}$ $e_g$ absorption peak, and the 2DES states are delayed from the $Ti^{4+}$ $t_{2g}$ peaks. This difference in excitation energies reflects the $Ti^{3+}$ $e_g$ character of the former *vs* $Ti^{4+}$ $t_{2g}$ character of the latter, for the full analysis see A. Chikina *et al*, ACS Nano **12** (2018) 7927.

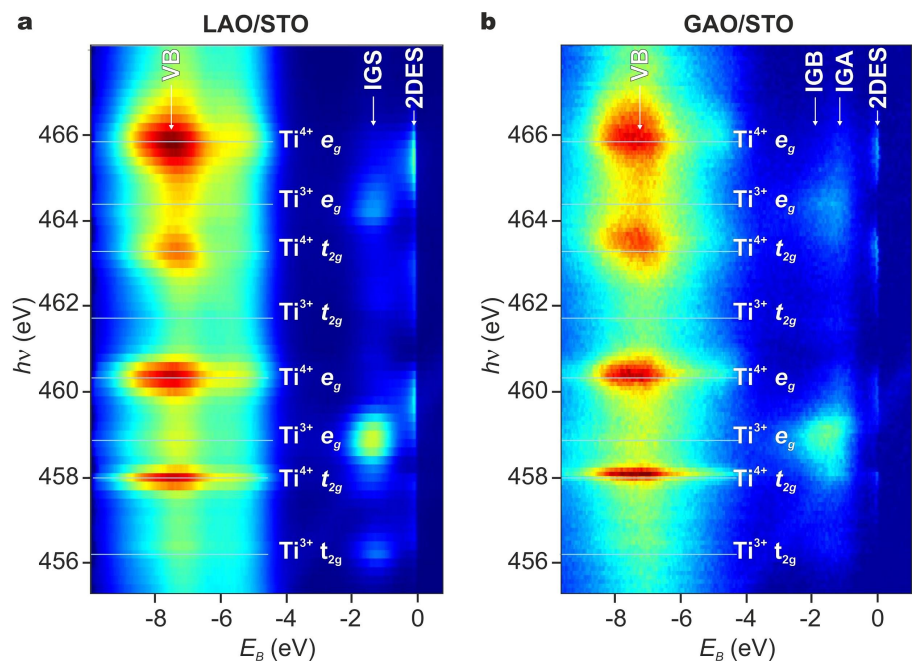

Fig. S1. ResPE maps for LAO/STO (a) and low-$n_s$ GAO/STO (b) samples. The latter shows two different IG states.

Remarkably, the ResPE map for GAO/STO shows two IG states, one at $E_B \sim 1.2$ eV (marked IGA) and another at -2.1 eV (marked IGB) (P. Schütz *et al*, Phys. Rev. B **96** (2017) 161409). The IGA spectral weight scales up under X-ray irradiation, and can be quenched by irradiation in oxygen atmosphere (A. Chikina *et al*, ACS Nano **12** (2018) 7927; V. N. Strocov *et al*, Phys. Rev. Materials

**3** (2019) 106001). The IGB, in contrast, is immune to the irradiation and oxygen atmosphere. While the IGA state is characteristic of the $V_O$s created by X-ray irradiation at bare STO surfaces and LAO/STO interfaces, the IGB one can be associated, tentatively, with fixed $V_O$s created during the sample growth in the top STO layer interfaced with GAO. The experimental ResPE map shows different excitation-energy dependence of the two states, identifying their different nature. Crystal-field calculations supporting this scenario of the formation of the IGA and IGB states were discussed in the Supplemental Materials of the above paper by P. Schütz *et al*.

# Effective-mass analysis

Experimental values of the effective mass $m^*/m_0$ ($m_0$ is the free-electron mass) for the $d_{xz}$ and $d_{yz}$ bands are compiled in the table below. They were calculated from the experimental values of $k_F$ indicated in Fig. 2 (determined from extremes of the gradient $dI/dk_x$ of the MDCs at $E_F$) and energies $E_{bot}$ of the $d_{xz}/d_{yz}$ band bottom. The uncertainty of $k_F$ and $E_{bot}$ has been determined from the scatter of their measurements through the experimental series.

| $m^*/m_0$ | LAO/STO | low-$n_s$ GAO/STO | high-$n_s$ GAO/STO |
|---|---|---|---|
| $d_{xz}$ | 0.72±0.32 | 0.32±0.22 | 0.40±0.23 |
| $d_{yz}$ | 11.5±2.5 | 6.0±1.2 | 9.8±1.3 |

# DFT calculations

In order to get more insight into the nature of electronic states at the GAO/STO interface, we performed DFT simulations for the supercell structures shown in Fig. S2 (a) that include an appropriate lattice matching between the two oxides and allow for modeling realistic concentrations of $V_O$s. Since the 2DES here is dominated by the Ti $3d$ electrons, the electronic correlations were taken into account using the GGA+$U$ corrections for these states (S. L. Dudarev *et al*, Phys. Rev. B **57** (1998) 1505). The GAO/STO interface was simulated by the superlattice (P. Schütz *et al*, Phys. Rev. B **96** (2017) 161409) where 4.5 u.c. of [001]-oriented STO were attached to five tetragonal and five octahedral Al layers of the $Al_3O_4$ spinel structure. One unit cell of GAO contains four layers of each type. We chose an in-plane (2×2) superlattice of the standard perovskite unit cell of STO to match the oxygen sublattices of the two materials. In order to take into account the defect-spinel structure of GAO where Al vacancies are distributed over the tetragonal and octahedral Al sites, we simulated a random distribution of cationic vacancies through the virtual crystal approximation (VCA) (L. Bellaiche and D. Vanderbilt, Phys. Rev. B **61** (2000) 7877) where the charge of each Al cation is shifted by 1/3 towards Mg leading to a nuclear charge Z=12+2/3. We consider a $V_O$ in the sublayer, Fig, S2 (a). In this case, the electrons donated by the $V_O$ are not fully localized.

The results of our DFT calculations provide important insights into the electronic structure of the spinel-perovskite GAO/STO interface. First of all, for most $V_O$ configurations the interface is conductive, and the **k**-resolved band structure in Fig. S2 (c) projected on the Ti sites indicates an inverted band ordering in GAO/STO with the $d_{xz}/d_{yz}$ orbitals lying well below $d_{xy}$. This is opposite to the situation typically observed in LAO/STO, Fig. S2 (d), and other oxide heterostructures (A. Chikina *et al*, ACS Nano **12** (2018) 7927). The hybridization between the $d_{xz}$ and $d_{yz}$ bands in GAO/STO, symmetry-forbidden in LAO/STO, results from the non-symmetric position of the $V_O$ in the supercell. The overestimate of $\mathbf{k}_F$ is an artifact of the limited supercell size in our calculations.

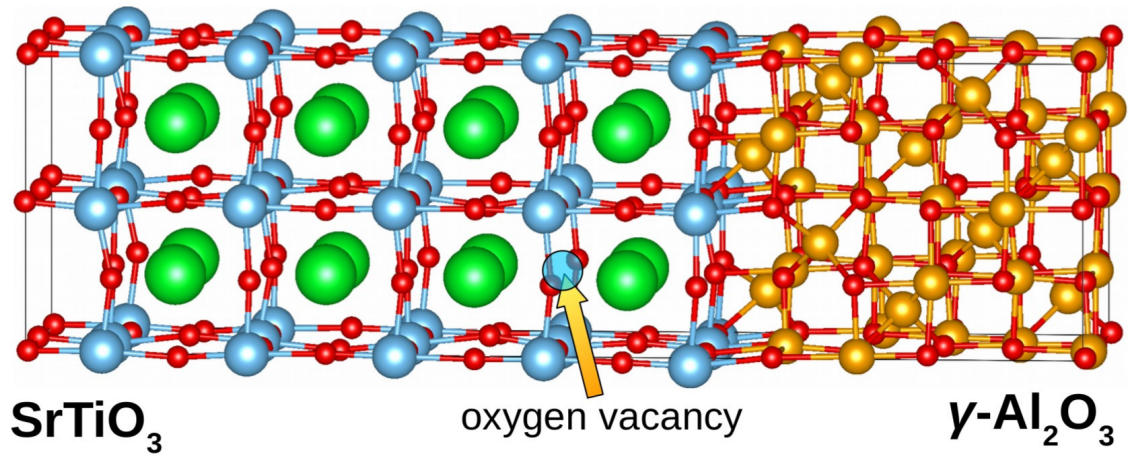

a) GAO/STO interface structure

SrTiO$_3$ — oxygen vacancy — γ-Al$_2$O$_3$

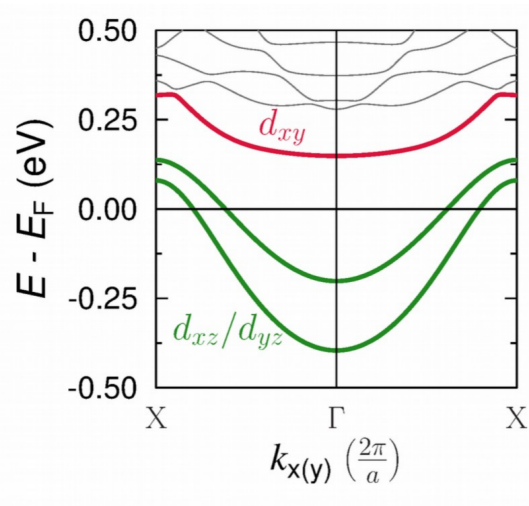

b) GAO/STO band structure

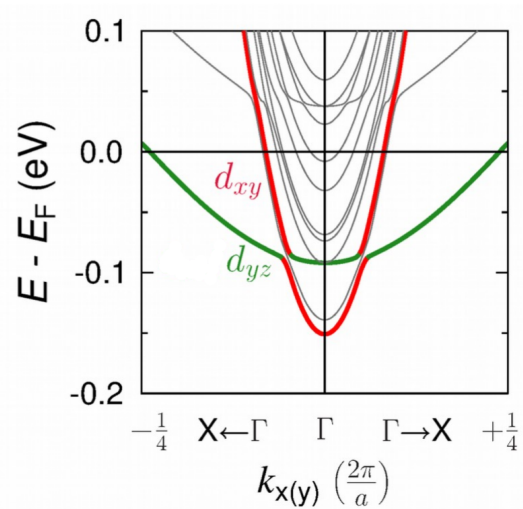

c) LAO/STO band structure

Fig. S2. Crystal and electronic structure of the oxygen-deficient GAO/STO interface. (a) Sketch of the GAO/STO supercell simulating the interface with a single V$_O$ (the arrow); (b,c) Calculated band structures for the oxygen-deficient GAO/STO and defect-free LAO/STO interfaces (the latter is adapted from J. Zabaleta *et al*, Phys. Rev. B **93** (2016) 235117) along the XΓX direction. The dominant orbital character is encoded by color: red ($d_{xy}$) and green ($d_{xz}$/$d_{yz}$).

# Experimental Fermi surface of GAO/STO

Another interesting peculiarity of GAO/STO seen in Fig. S3 (a), absent for LAO/STO, is notable streaks of the Fermi intensity stretching along the ΓX directions across the BZs. This peculiarity should manifest a lateral disorder in GAO/STO along the (100) crystallographic directions. Such a disorder can manifest formation of a rectangular-pattern, as schematized in Fig. S3 (c). The origin of this disorder is not yet clear, but potential candidates could be (1) linear clusters of the interfacial $V_O$s in GAO/STO, with their linear arrangement being slightly energetically preferred (F. Cordero, Phys. Rev. B **76** (2007) 172106; D. D. Cuong *et al*, Phys. Rev. Lett. **98** (2007) 115503), or (2) rectangular domains, whose dimensions are comparable with the ARPES coherence length, although our scanning-SQUID measurements on similarly prepared samples have only found micron-size domains (D. Christensen *et al*, Phys. Rev. Applied **9** (2018) 054004) typical of the STO-based systems (Frenkel *et al*, Nat. Mater. 16 (2017) 1203). We note that similar streaks in the FS map were observed at bare $BaTiO_3$ surfaces and interpreted in terms of the Wannier-Stark electron localization in a strong electric field between the ferroelectric domains of $BaTiO_3$ (S. Muff *et al*, Phys. Rev. B **98** (2018) 045132). Such a strong electric field is however absent at the GAO/STO interface.

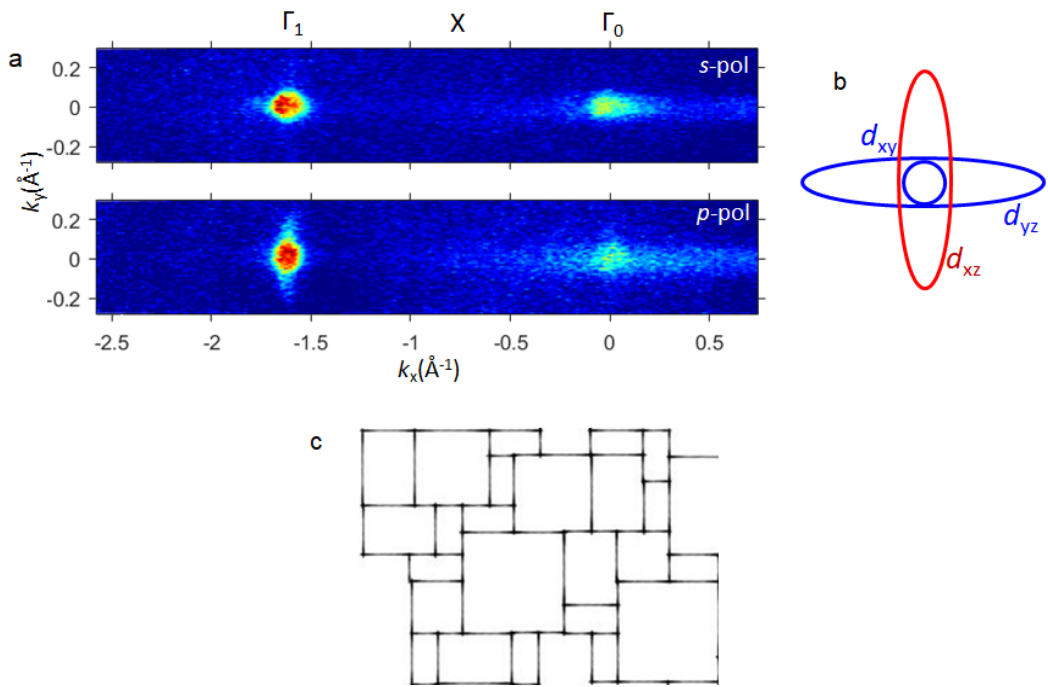

Fig. S3. (a) FS map of GAO/STO acquired at the Ti $L_3$-edge ($h\nu$=460.2 eV). *s*- and *p*-polarization of incident X-rays selects electron states of different symmetry as schematized in (b); (c) A random-rectangle disorder pattern explaining the Fermi-intensity streaks along the ΓX direction.